\documentclass[setspace,amsmath,amssymb]{article}
\usepackage[all]{xy}
\usepackage{graphicx}
\usepackage{setspace}
\usepackage{amsmath}
\usepackage{amssymb}
\usepackage{rsfs}
 
\newcommand{\ra}{\rangle}

\newcommand{\be}{\begin{equation}}
\newcommand{\ee}{\end{equation}}
\newcommand{\bea}{\begin{eqnarray}}
\newcommand{\eea}{\end{eqnarray}}

\begin{document}
\begin{titlepage}

\begin{flushright}
\today
\end{flushright}

\vspace{1in}

\begin{center}

{\bf A remark on quantum measuring systems and the holographic principle}

\vspace{1in}

\normalsize

{Eiji Konishi\footnote{E-mail address: konishi.eiji.27c@kyoto-u.jp}}

\normalsize
\vspace{.5in}

{\it Graduate School of Human and Environmental Studies,\\
 Kyoto University, Kyoto 606-8501, Japan}
\end{center}

\vspace{1in}

\baselineskip=24pt
\begin{abstract}
It is a sort of ultimate question to examine the continuity of a quantum measurement subject theoretically and has not yet been resolved within a scientific framework.
In this article, we approach this question and argue that the continuity of a quantum measurement subject follows as a fundamental consequence of the holographic principle after the classicalization of the quantum state of the bulk space.

\end{abstract}

\vspace{.1in}


\vspace{.6in}


\end{titlepage}


\section{Introduction}

It is a sort of ultimate question to examine the continuity of a quantum measurement subject theoretically and has not yet been resolved within a scientific framework.
The purpose of this article is to approach this question based on the author's previous work on the holographic principle and quantum measurement theory and to argue that the continuity of a quantum measurement subject follows as a fundamental consequence of the holographic principle \cite{Hol1,Hol2,Hol3} after the classicalization of the quantum state of the bulk space.

In this article, all arguments are based on the following two equations in the three-dimensional anti-de Sitter spacetime/two-dimensional conformal field theory (AdS$_3$/CFT$_2$) correspondence, which is a known example of the holographic principle \cite{AdSCFT1,AdSCFT4}:
\begin{enumerate}
\item[I.] Quantum ground state.
Namely, the low-energy limit of bulk quantum gravity before the classicalization:
\begin{equation}
\left(\widehat{H}_{\rm CFT}-E_{\rm gnd}\right)|\psi\ra=0\;.
\end{equation}
Here, $\widehat{H}_{\rm CFT}$ is the Hamiltonian operator of the boundary CFT$_2$ at the strong-coupling limit, and $E_{\rm gnd}$ is its ground state energy (i.e., a negative Casimir energy).

\item[II.] Euclidian bulk action.
Namely, the holographic principle after the classicalization of the quantum ground state $|\psi\ra$ \cite{EPL1,EPL2,JHAP1}:
\begin{equation}
I_{\rm bulk}[|\psi\ra,\gamma_\tau]=-\hbar H[|\psi\ra]+S_E[\gamma_\tau]\;.\label{eq:Ibulk}
\end{equation}
Here, $\gamma_\tau$ is the trajectory of a non-relativistic free particle in the Euclidean bulk spacetime, $S_E$ is the Euclidean action of this particle, and $H$ is the Shannon entropy (i.e., the von Neumann entropy of the diagonal classicalized quantum state) in nats.

\end{enumerate}

Throughout this article, {\it quantum measurement} refers to a process in the ensemble interpretation of quantum mechanics \cite{dEspagnat}.

We organize the rest of this article as follows.

In Sec. 2, we summarize the results obtained so far in the context of quantum measurements \cite{JHAP1,JHAP4}.
In Sec. 2.1, we argue that the theory is fundamentally formulated in the Euclidean regime by the Euclidean bulk action (\ref{eq:Ibulk}).
In Sec. 2.2, we transfer the time evolution from imaginary time to real time by the inverse Wick rotation.

In Sec. 3, we present the main argument of this article.
In Sec. 3.1, we describe the physical transfer of a quantum measurement subject accompanying the extinction of the subject of a quantum measuring system.
In Sec. 3.2, we manifest the continuity of the quantum measurement subject.

In Sec. 4, we summarize the subjects of quantum measurement in imaginary time and real time, and their relations.

\section{Quantum measurements}

\subsection{Euclidean regime}

On the codimension-one boundary spacetime of AdS$_3$ spacetime, CFT$_2$ at the strong-coupling limit is in the quantum ground state $|\psi\ra$, and we construct the quantum state of the bulk space by classicalizing the multi-scale entanglement renormalization ansatz (MERA) of $|\psi\ra$ \cite{EPL1,EPL2,Vidal1,Vidal2,Swingle1}.
Here, {\it classicalization} means the exact and complete loss of quantum coherence in the Hilbert space of each qubit constituting the MERA by application of the superselection rule, where the superselection rule operator is the third Pauli matrix $\sigma_3$.
Here, at each site (disentangler) of the bulk space, the statistical mixture of eigenstates of a spin operator gives a $\ln 2$ nat increment of the von Neumann entropy (the measurement entropy) of the quantum state of the classicalized bulk space \cite{EPL2}; the spin operator is locally defined at each site \cite{EPL2,JHAP1}.
The total amount of the measurement entropy in bits is the number of sites in the classicalized bulk space \cite{EPL2}.

The division of the off-shell value of the Euclidean action of a free particle moving in the bulk space in imaginary time by the Dirac constant $\hbar$ and the bit factor $b=\ln 2$ means the information of the sequence of spin eigenstates in the mixture of spin eigenstates \cite{JHAP1}; this mixture is locally owned by the bulk space itself at each site \cite{EPL2,JHAP1}.
Namely, the process of this imaginary-time motion is the sequence of readouts of spin events from the mixture of spin events on the boundary spacetime \cite{JHAP1}.
Thus, due to the imaginary-time motion of free particles in bulk space, in imaginary time, the quantum spin measurement process of the quantum ground state $|\psi\ra$ on the codimension-one boundary spacetime is realized in the exact and complete form as follows:
\begin{center}
Quantum ground state $|\psi\ra$ $\longrightarrow$ Mixture of spin eigenstates: classicalization $\longrightarrow$ Mixture of sequences of spin eigenstates.
\end{center}
Here, the mixture of sequences of spin eigenstates means the sequence of readouts of spin events due to the Euclidean action of free particles in the imaginary-time motion in the bulk space.

Here, we stress that {\it the bulk space is the measurement entropy of the classicalized quantum ground state $|\psi\ra$} \cite{EPL2}.

\subsection{Lorentzian regime}

The real-valued classical probability in the mixture of sequences of spin events, which results from the quantum spin measurement process of the quantum ground state $|\psi\ra$, becomes a complex-valued quantum probability amplitude by the inverse Wick rotation.
Thus, in real time, for a quantum mechanical event to occur, a quantum measurement by a quantum measuring system must be done.

Here, the {\it quantum measuring system} is a macroscopic composite system \cite{Araki,EPL3}, which, in real time, can change the quantum probability amplitude of a measured system in the bulk spacetime, namely, the quantum pure state, to its classical mixed state (i.e., its diagonal part as a mixture of eigenstates) with respect to the discrete measured observable by the interaction with the measured system (non-selective measurement) and can change the classical mixed state to a classical pure state (event reading).

Concretely, subject to the {\it orbital superselection rule} (which restricts the complete set of orbital observables of the center of mass of a macroscopic system to an Abelian set of simultaneously measurable classical observables \cite{Neumann}), the Hamiltonian (which gives rise to the measurement process by a quantum measuring system), except for the free parts, consists of the von Neumann-type interaction with the center of mass of a macroscopic system \cite{Araki,EPL3}.

Here, there are two kinds of the macroscopic system.
In the realization of non-selective measurement \cite{Araki}, the system is a macroscopic detector, whose center of mass interacts with the measured system.
In the realization of event reading \cite{EPL3}, the system is a macroscopic Bose--Einstein condensate with spontaneously broken spatial translational symmetry, whose center of mass interacts with a quantum mechanical system having a discrete meter variable for the event reading.
The orbital superselection rule gives rise to the classical mixed state of Planck cells in the phase space as the quantum state of a macroscopic system \cite{Neumann}.

\section{Continuity of the subject of quantum measurement}

In the following, {\it we define the presence or absence of the subject by the presence or absence of event reading by a quantum measurement, respectively}.

\subsection{Description}

Here, we invoke the idea of integrated information theory (IIT) proposed by Tononi \cite{IIT0}.
In the author's opinion, the most important idea in IIT is to quantify the level of the subject of a quantum measuring system, and then to abandon the concept of {\it self-identity}, which means the maintenance of the consistency of the subject of a quantum measuring system.

We state this in a more concrete form.\footnote{For simplicity, we explain IIT 2.0 \cite{IIT1,IIT2}.
The original paper of IIT 3.0 (resp., IIT 4.0) is Ref. \cite{IIT3} (resp., Ref. \cite{IIT4}).}
First, we define the {\it integrated information} for the event reading quantum mechanical system as a discrete dynamical system.
We define the integrated information for the dynamical causal structure in this system and the state of this system---namely, the event of this system read out by itself---at an instance (say, $t=0$) by the minimum amount of lost information in the {\it a posteriori} state distribution of this system just before the readout instance (say, $t=-1$) by the possible cuts of the dynamical causal relations between elements of this dynamical system \cite{IIT1}.
Second, we quantify the {\it level of the subject of a quantum measuring system} by the capacity of integrated information owned by the subject of this quantum measuring system \cite{IIT0}.
Based on this quantification of the level, we can describe the process of complete loss of the level of the subject of a quantum measuring system, in which the concept of self-identity collapses, as a physical process in real time.

In real time, the physical process of the {\it extinction of the subject of a quantum measuring system} consists of two steps:
\begin{enumerate}
\item[1.] The complete loss of the capacity for integrated information owned by the subject of the quantum measuring system, namely, the level of the subject of the quantum measuring system.
Here, the self-identity applied to the subject of the quantum measuring system is resolved.
\item[2.] The physical extinction of the quantum mechanical events themselves.
Furthermore, the real-time evolution of the quantum measuring system becomes a unitary process.
\end{enumerate}

However, equivalently corresponding to the unitary real-time evolution of free particles in a quantum pure state in the bulk space, in imaginary-time evolution, quantum spin measurements continue; namely, spin events continue to occur.
Here, the completely self-identical {\it hologram}, namely, the quantum ground state $|\psi\ra$ on the codimension-one boundary of the bulk space, is the unique subject of quantum measurements.

Specifically, we can conclude that the physical transfer of the subject of quantum measurement accompanying the extinction of the subject of a quantum measuring system is the following.
\begin{enumerate}
\item[i.] The dimensional descent from the quantum state of the quantum measuring system as a macroscopic composite system in the bulk spacetime to the classicalized quantum ground state $|\psi\ra$ of the hologram on the codimension-one boundary spacetime through the extinction of the subject of the quantum measuring system.
\end{enumerate}
and
\begin{enumerate}
\item[ii.] The transfer of the time evolution equivalently from real time to imaginary time by the Wick rotation at the time i.
\end{enumerate}

Here, the information acquired by the classicalized quantum ground state $|\psi\ra$, as the unique subject of quantum measurement, in the imaginary-time evolution is the information generated by the imaginary-time motion of all free particles in the bulk space \cite{JHAP4}.

\subsection{Manifestation}

We denote by $\Phi_{\rm max}$ the capacity of integrated information owned by the subject of a quantum measuring system.
By using the notation in Ref. \cite{JHAP4}, we can express process 1 as the extinction of the experience \cite{IIT0} and process 2 as the extinction of the existence in the Lorentzian regime \cite{JHAP4}, along with their respective reverse processes 4 and 3, by the following diagram:
\begin{equation}
{{\xymatrix{
\Phi_{\rm max}>0 \ar[d]\ar[r]^{1}&\Phi_{\rm max}=0 \ar[dl]\ar@{-->}[d]\ar[r]^{4}\ar[dr]&\Phi_{\rm max}>0\;.\ar[d]\\
S_{\rm v.N.}>0\;,\ I\ge 0\ar[rd]^{2}\ar@{-->}[r]^{2}&S_{\rm v.N.}=0\;,\ I=0\ar@{.>}[d]^{\rm a.c.}\ar@{-->}[r]^{3}&S_{\rm v.N.}>0\;,\ I\ge 0 \\
&S_{\rm v.N.}=A_{\rm TN}\;,\ {\displaystyle{I=\frac{S_E}{\hbar b}}}\ar[ur]^{3}&}}}
\end{equation}
Here, $S_{\rm v.N.}$ is the von Neumann entropy, in bits, of the quantum state of the quantum measuring system in the second line (the Lorentzian regime) and the classicalized hologram in the third line (the Euclidean regime), $I$ is the information in bits acquired by the quantum measuring system in the second line and the classicalized hologram in the third line, and $A_{\rm TN}$ is the number of sites in the classicalized bulk space.
The dotted arrow indicates the analytic continuation (a.c.) from the Lorentzian regime to the Euclidean regime \cite{JHAP4}.

This diagram manifests the continuity of the subject of quantum measurement.
Indeed, by taking out the subject of quantum measurement from this diagram, we obtain
\begin{equation}
{{\xymatrix{
\Phi_{\rm max}>0 \ar[d]\ar[r]^{1}&\Phi_{\rm max}=0 \ar[dl]\ar[r]^{4}\ar[dr]&\Phi_{\rm max}>0\;.\ar[d]\\
S_{\rm v.N.}>0\;,\ I\ge 0\ar[rd]^{2}&&S_{\rm v.N.}>0\;,\ I\ge 0 \\
&S_{\rm v.N.}=A_{\rm TN}\;,\ {\displaystyle{I=\frac{S_E}{\hbar b}}}\ar[ur]^{3}&}}}
\end{equation}
Here, processes 2 and 3 are accompanied by the Wick rotation and the inverse Wick rotation, respectively.

\section{Summary}

We summarize this article in the following four points.

In imaginary time (the Euclidean regime), the subject of quantum measurement is the classicalized hologram on the boundary spacetime only.

In real time (the Lorentzian regime), the subjects of quantum measurement are the quantum measuring systems in the bulk spacetime only.

The quantum measurement process in the Euclidean regime is equivalent to path integral unitary quantum mechanics in the Lorentzian regime via the inverse Wick rotation.

By the extinction of the subject of a quantum measuring system, the Lorentzian regime becomes equivalent to the Euclidean regime via the Wick rotation.


\begin{thebibliography}{99}
\bibitem{Hol1}G. 't Hooft, arXiv:gr-qc/9310026.
\bibitem{Hol2}L. Susskind,
J. Math. Phys. {\bf 36}, 6377 (1995).
\bibitem{Hol3}R. Bousso,
Rev. Mod. Phys. {\bf 74}, 825 (2002).
\bibitem{AdSCFT1}J. M. Maldacena, 
Adv. Theor. Math. Phys. {\bf 2}, 231 (1998).
\bibitem{AdSCFT4}O. Aharony, S. S. Gubser, J. M. Maldacena, H. Ooguri and Y. Oz,
 Phys. Rep. {\bf 323}, 183 (2000).
\bibitem{EPL1}E. Konishi, 
EPL {\bf 129}, 11006 (2020).
\bibitem{EPL2}E. Konishi,
 EPL {\bf 132}, 59901 (2020), arXiv:1903.11244 [quant-ph].
\bibitem{JHAP1}E. Konishi,
JHAP {\bf 1}, (1) 47-56 (2021).
\bibitem{dEspagnat}B. d'Espagnat, {\it Conceptual Foundations of Quantum Mechanics}, 2nd edn. (W. A. Benjamin, Reading, Massachusetts, 1976).
\bibitem{JHAP4}E. Konishi,
JHAP {\bf 3}, (1) 31-38 (2023).
\bibitem{Vidal1}G. Vidal, 
Phys. Rev. Lett. {\bf 99}, 220405 (2007).
\bibitem{Vidal2}G. Vidal, 
Phys. Rev. Lett. {\bf 101}, 110501 (2008).
\bibitem{Swingle1}B. Swingle, 
Phys. Rev. D {\bf 86}, 065007 (2012).
\bibitem{Araki}H. Araki,
Prog. Theor. Phys. {\bf 64}, 719 (1980).
\bibitem{EPL3}E. Konishi,
EPL {\bf 136}, 10004 (2021).
\bibitem{Neumann}J. von Neumann, {\it Mathematical Foundations of Quantum Mechanics} (Princeton University Press, Princeton, NJ, 1955).
\bibitem{IIT0}G. Tononi, M. Boly, M. Massimini and C. Koch, 
Nat. Rev. Neurosci. {\bf 17}, 450 (2016). 
\bibitem{IIT1}D. Balduzzi and G. Tononi,
PLoS Comput. Biol. {\bf 4}, e1000091 (2008).
\bibitem{IIT2}D. Balduzzi and G. Tononi,
PLoS Comput. Biol. {\bf 5}, e1000462 (2009).
\bibitem{IIT3}M. Oizumi, L. Albantakis and G. Tononi,
PLoS Comput. Biol. {\bf 10}, e1003588 (2014).
\bibitem{IIT4}L. Albantakis, L. Barbosa, G. Findlay, M. Grasso, A. M. Haun, W. Marshall, et al.,
PLoS Comput. Biol. {\bf 19}, e1011465 (2023).
\end{thebibliography}
\end{document}